\documentstyle[12pt]{article}

\begin{document}

\begin{center}
{\Large Hadron Helicity Violation in Exclusive Processes:\\
Quantitative Calculations in Leading Order QCD}

\vspace{1.5cm}

{\bf Thierry Gousset}

\medskip
{\it CEA, Service de Physique Nucl\'eaire/DAPNIA,\\
CE Saclay, F91191 Gif, France\\
and\\ Centre de Physique Th\'eorique,\\
Ecole Polytechnique, F91128 Palaiseau, France}

\bigskip
{\bf Bernard Pire}

\medskip
{\it Centre de Physique Th\'eorique,\\
Ecole Polytechnique, F91128 Palaiseau, France}

\bigskip
{\bf John P. Ralston}

\medskip
{\it Department of Physics and Astronomy\\ 
and\\ Kansas Institute for Theoretical and Computational Science,\\
University of Kansas, Lawrence, Kansas 66045 USA}

\end{center}

\vspace{2cm}

\begin{abstract} 
We study a new mechanism for hadronic helicity flip
in high energy hard exclusive reactions. The mechanism
proceeds in the limit of perfect chiral symmetry, namely without any need to
flip a quark helicity. The fundamental feature of the new mechanism is the
breaking of rotational symmetry of the hard collision by a scattering plane in
processes involving independent quark scattering. We show that in the impulse
approximation there is no evidence for of the helicity violating process 
as the energy or momentum transfer $Q^2$ is increased over the region 
$1\,GeV^2<Q^2<100\,GeV^2$. In the asymptotic region $Q^2> 1000$ GeV$^2$,
a saddle point approximation with doubly logarithmic accuracy yields 
suppression by a fraction of power of $Q^2$. 
``Chirally--odd" exclusive wave functions which carry non--zero orbital 
angular momentum and yet are leading order in the high energy limit, play 
an important role. 
\end{abstract}

\newpage


\section{Introduction}

The theory of hard elastic scattering in Quantum
Chromodynamics (QCD) has evolved considerably over many years of work.
Currently there exist two self-consistent perturbative descriptions, each 
with a specific factorization method for separating the hard scattering 
from non-perturbative wave functions. A well-known procedure using the
``quark-counting'' diagrams has been given by Lepage and Brodsky~\cite{bro1}. 
A consequence, and direct test, of the factorization defining this mechanism 
is the hadron helicity conservation rule~\cite{bro2}
\begin{equation} \label{hel}
\lambda_A+ \lambda_B =  \lambda_C  + \lambda_D\; ,
\end{equation}
where the $\lambda_j$'s are the helicities of the participating hadrons in 
the reaction $A+B\rightarrow C+D$.  
The fact that this rule is badly violated in almost every case tested 
suggests two alternatives. One possibility, advocated by Isgur and 
Llewellyn-Smith~\cite{isg}, is that the energy and momentum transfer ($Q^2$) 
in the data is not large enough for the formalism to apply. However, the data
also shows behavior close to the model's predicted power dependence on $Q^2$,
indicating that hard scattering of a few pointlike quarks is being
observed. The apparent contradiction has led to much discussion, and has even
caused some authors to suggest that perturbative QCD itself might be wrong.

The second alternative is that another power behaved process causing
helicity flip is present.  In fact the ``independent scattering'' subprocess,
introduced by Landshoff~\cite{lan}, is actually the leading process at very 
high energies~\cite{mue}. But it has been assumed that hadron helicity 
conservation would be the same in the independent scattering process as in 
the quark-counting one, since both involve exchange of hard gluons with 
large $Q^2$. In general, terms proportional to a quark mass, $m_q$, for 
example, have been understood to cause helicity flip in either model, but 
with amplitude suppressed by a power of $m_q/Q$ relative to the leading 
term. Such terms seem to be quite small and are probably not a believable 
explanation of the persistent pattern of large violations of the helicity 
conservation rule.

Here we show that the independent scattering mechanism predicts high-energy 
helicity {\it non-conservation} at a much higher rate. The calculations in 
momentum space are sufficiently complicated that this phenomenon has been 
overlooked for almost twenty years. Adopting a transverse position space 
formalism introduced by Botts and Sterman~\cite{bot}, we show that the 
details rest on non-perturbative wave functions that should be {\it measured} 
rather than calculated. These wave functions measure non--zero orbital angular 
momentum not taken into account by short distance expansions. We argue that 
the novel factorization properties of independent scattering processes 
cannot practically be reduced to the same ingredients used in the quark 
counting scattering. In any case, it is not necessary to flip a quark 
helicity: the new mechanism proceeds unimpeded in the limit of arbitrarily 
small quark mass and perfect chiral symmetry in the hard scattering.

This paper is organized as follows. In section~2 we review the derivation 
of helicity conservation in genuine short distance processes.. In section~3, 
we present the independent scattering mechanism with special emphasis on 
non-zero orbital angular momentum wave functions. We compute the 
contribution of these components to a helicity conserving reaction in 
section~4, then to a helicity violating reaction in section~5. These two 
sections contain our most important results at asymptotic and at accessible 
energies. Section~6 is an experimental outlook.



\section{Helicity conservation in short distance dominated processes}

First we review the conventional derivation of hadron helicity 
conservation~\cite{bro2}. The quark-counting factorization introduces the 
distribution amplitude $\varphi(Q^2,x)$~\cite{bro3}, an integral
over the transverse momentum variables of the wave function for quarks to be
found carrying momentum fraction $x$ in the hadron.\footnote{Our
convention does not include a logarithmically varying factor of $Q^2$
introduced for renormalization group analysis in Ref.~\cite{bro1,bro3}. Either 
convention can be used without affecting our argument.}  
For simplicity of presentation we specialize to a single pair of quarks, the 
meson case. Let $\psi({\bf k}_T,x)$ be a light cone wave 
function to find quarks with relative transverse momentum
${\bf k}_T$ and light cone momentum fraction $x$. In terms of the Fourier 
conjugate transverse space variable ${\bf b}_T$ separating the quarks, then
\begin{eqnarray}
&&\varphi(Q^2,x)=\int\limits^Q_0d^2{\bf k}_T \;\psi({\bf k}_T,x)
\nonumber\\ \label{orbital}
&&=\int\limits^Q_0d^2{\bf k}_T\int\limits^\infty_0d^2{\bf b}_T
e^{i{\bf b}_T\cdot{\bf k}_T}
\sum\limits_m e^{im\varphi}\tilde\psi_m(\mid{\bf b}_T\mid,x) 
\end{eqnarray}
In the second line we have expanded the wave function 
$\tilde\psi({\bf b}_T, x)$ to exhibit the $SO(2)$ orbital angular momentum 
eigenvalues $m$: 
a complete set for the (nearly on-shell) quarks consists of the $z$-axis 
orbital angular momentum, the energy, and the $z$-momentum.  Now suppose the
distribution amplitude $\varphi(Q^2,x)$ is assumed to be a good description of a
process. Then, whatever the angular momentum of the wave function, evaluating 
the integrals reveals the $m = 0$ element is the sole surviving term in 
Eq.~(\ref{orbital}):
\begin{equation}
\varphi(Q^2,x)=2\pi\int\limits^\infty_0d\mid {\bf b}_T\mid QJ_1(\mid
{\bf b}_T\mid Q)\tilde\psi_0(\mid{\bf b}_T\mid,x)
\end{equation}
This shows that use of $\varphi(Q^2,x)$ imposes two things: as
$Q^2\rightarrow\infty$ the scattering region is both ``small'', since the 
region ${\bf b}_T^2 <1/Q^2$ dominates in the Bessel function, and the 
scattering is ``round'', {\it i.e.} cylindrically symmetric (Fig.~1). In the 
absence of orbital angular momentum, the hadron helicity becomes the sum 
of the quark helicities.  The quark helicities being conserved 
at leading order, the total hadron helicities are conserved.  The hadron 
helicity conservation rule (1) therefore represents an exact {\it symmetry} 
of the quark-counting factorization.  A crucial question is: does this 
symmetry of the model represent a  property of the entire perturbative 
theory. Or, can we simply assume ``s--wave'' $SO(2)$ wave functions 
to be the main contribution as in a non--relativistic picture?

The answer to both questions is {\it no}. 
In general, quark wave functions themselves are not particularly restricted 
in orbital angular momentum content, even in the high energy limit. For 
example, in the pion (pseudoscalar meson) case the light-cone wave function 
may be expanded on four Dirac tensors  
allowed by parity symmetry as
\begin{equation}
\label{pion}
\tilde\psi_{\alpha\beta}(x,{\bf b}_T;p^\mu)=
\left\{ A_\pi /\!\!\! p\gamma_5+B_\pi\gamma_5 +C_\pi /\!\!\!b_T\gamma_5
+D_\pi[/\!\!\!p,/\!\!\!b_T]\gamma_5\right\}_{\alpha\beta},
\end{equation}
\noindent
where $A_\pi$-$D_\pi$ are functions of the light cone fraction $x$, the 
transverse separation ${\bf b}_T$ and the total four momentum of the meson 
$p^\mu$. For the moment we do not discuss dependence on gauge fixing and 
a path-ordered exponential of the gauge potential used to 
make the wave function gauge invariant.
The $D_\pi$-term carries one unit of orbital angular momentum and yet scales 
with the same power of the ``big'' momentum $p^+$ as the $A_\pi$--term, which 
is s-wave. Since the $D_\pi$ term has a $\vec b_T$ factor, which can be
written in terms of $b_{T,1}\pm ib_{T,2}$ this term represents one unit of
orbital angular mometum. In terms of power counting, then, the $m=0$ 
and $m\neq 0$ amplitudes can be equally large. We also note that 
wave functions are not objects to be derived in perturbation
theory, but instead represent the long-time, non-perturbative time evolution
proceeding inside a hadron. The non-perturbative Hamiltonian of QCD does not
conserve spin and orbital angular momentum separately, but instead generates
mixing between orbital and spin angular momentum. Finally, there is no simple
relation between ``s--wave'' non--relativistic models of constituent quarks, 
and the pointlike quarks resolved in large--$Q^2$ collisions, so no statement 
can reliably be made about quark angular momentum content of hadrons. Thus 
{\it if} a non-zero orbital angular momentum component somehow enters the 
hard scattering --- and this is a crucial point --- {\it then} the long-time 
evolution before or after the scattering can convert this angular momentum 
into the observed hadron spin. It is not necessary to flip a quark spin in 
the hard interaction, because the asymptotic hadron spin fails to equal the 
sum of the quark spins. Such a mechanism is totally consistent with the 
impulse approximation of perturbative QCD.

The challenge in high energy hadron scattering is therefore to find those 
large $Q^2$ process(es) in which non-zero orbital angular momentum enters, or 
in other words, to find those which are not ``round''. It turns out that 
in any treatment relevant to current energies the independent scattering 
process is not ``round'' but instead ``flat''~(Fig.~1).
The subprocess is highly asymmetric, showing an extreme dependence on the 
scattering plane. Rather than disappearing in the high energy limit, 
the dynamics of this process {\it increases its asymmetry over a range of} 
$Q^2$.

The origin of the asymmetry is kinematic. Let $\eta$ be the direction
perpendicular to the scattering plane and $\xi$ be a vector in the scattering
plane. In the independent scattering mechanism, let two (or more) uncorrelated 
scattering planes be separated at the collision point by a transverse 
out--of--plane distance $b_{\eta}$. The out--going beams of quarks must 
coincide in direction well enough to make hadrons in the final state. 
Conserving 4-momenta, for each pair of scatterings there are three delta 
functions of large momenta (scaling like $p^+$), and one delta function of 
the out-of plane transverse momentum, of the order of $1/<b_{\eta}^2>^{1/2}$, 
where $< >$ indicates a typical expectation value in the state. On--shell 
quark--quark scattering is dimensionless and scale invariant (up to 
logarithmic corrections) in QCD. The overall amplitude therefore scales like 
the product of these factors, namely like $<b_{\eta}^2>^{1/2} (Q^2)^{-3/2}$. 
The $Q^2$ power-counting of this argument is quite well-known; what we 
emphasise here is the role of the scattering plane, namely the breaking of 
rotational symmetry with the out-of-plane direction $\eta$. It is as if 
hadrons ``flatten'' under impact in the in-plane direction $\xi$, forming a 
cigar-shaped hard scattering region~\cite{ral}. This fact is very hard to see 
in covariant perturbation theory in momentum space, explaining why it has 
generally been overlooked.


The kinematic violation of hadron helicity conservation by independent
scattering raises several new questions. It is clear that the usual association
of leading--twist (short--distance) and large $Q^2$ either breaks down or
hinges on delicate dynamical details. Our approach will exploit the fact that
leading approximations to any kinematically distinct amplitude are always
perturbatively calculable. For example, higher order corrections of
non--leading twist type in the distibution amplitude formalism cannot violate
the hadron helicity conservation symmetry and will not affect our approach. The
first non--vanishing contributions to helicity violating amplitudes involve
extra partons in the short distance formalism.  A gluon embedded in the hard 
scattering, for example, could transfer spin to an outgoing hadron. We need not 
consider such processes, because, as discussed later, they are subleading 
by a power of $Q^2$ and perturbatively small since such gluons are ``hard''.
It remains to be shown, of course, that helicity violation from independent 
scattering is not suppressed by the same order. That
is the main technical task of this paper.


\section{Independent scattering: formalism}

\subsection{Kinematical analysis}
Botts and Sterman have considered~\cite{bot} the generic ``elastic'' reaction 
$M_1+M_2\rightarrow M_3+M_4$, where $M_i$'s are light pseudoscalar mesons, 
at high energy $\sqrt{s}$ and large angle center of mass scattering angle 
$\theta=\arccos(1+2t/s)$. 
Their study has shown that the reaction is dominated by the 2 independent 
scatterings of the valence constituents with a kinematical configuration 
depicted in Fig.~2. One has two scattering planes separated at the collision 
point by a transverse distance $b$.

To make it clear, let us consider a light cone basis $(v_i,v_i',\xi_i,\eta)$ 
attached to each meson $M_i$, and chosen so that, in the center of mass frame
where $\widehat{p_1}=\hat{3}$ and 
$\widehat{p_3}=\cos\theta\,\hat{3}+\sin\theta\,\hat{1}$,
\[
\begin{array}{ll}
v_1=v_2'={1\over\sqrt{2}}(\hat{0}+\hat{3});&
v_1'=v_2={1\over\sqrt{2}}(\hat{0}-\hat{3});\\
\noalign{\vskip2mm}
\xi_1=\xi_2=\hat{1};&
\eta=\hat{2};\\
\noalign{\vskip2mm}
v_3=v_4'={1\over\sqrt{2}}(\hat{0}+\sin\theta\,\hat{1}+\cos\theta\,\hat{3});
\hspace{10mm}&
v_3'=v_4={1\over\sqrt{2}}(\hat{0}-\sin\theta\,\hat{1}-\cos\theta\,\hat{3});\\
\noalign{\vskip2mm}
\xi_3=\xi_4=\cos\theta\,\hat{1}-\sin\theta\,\hat{3}.
\end{array}
\]
\noindent
In the cm frame, one has $p_i=Qv_i$ ($Q=\sqrt{s/2}$). Each Bethe-Salpeter 
amplitude $X_i$ 
\[
X_i(k,p_i)=\int {d^4k\over(2\pi)^4}e^{ik.y}
\langle0|T(\psi_{\alpha}(y)\bar{\psi}_{\beta}(0))|\pi(p_i)\rangle,
\]
is assumed to 
eliminate $q\bar{q}$ configuration with relative $O(Q)$ transverse momentum 
(along $\xi_i$ and $\eta$). The relative minus momentum (along $v_i'$) is of
$O(M^2/Q)$. Then, momentum transfers in $H$ or $H'$ 
are dominated by large invariants built with $x_iQ$-terms and a first 
approximation is to neglect all but these components 
of the quark or antiquark momenta, in the hard amplitudes $H$ and $H'$. This 
is the impulse approximation.


A second observation follows from kinematics; although momentum 
conservation at the hard scattering $H$ relates the internal momentum 
dependence of the $X$'s, a variation of one such momentum $k_i$ in its 
$\xi_i$ or $v_i'$ direction induces negligible modifications in the three 
other $X$'s. Consequently, momentum components of $k_i$ along $\xi_i$ 
or $v'_i$ only appear as a relevant variable in the wave function $X_i$
and integrals over these components can be carried out. The components along 
$\eta$, denoted $l_i$, represent transverse momentum out of the scattering
plane and do not share the same property. Thus, the vertex delta 
function may be reexpressed as
\[
\delta^4(k_1+k_2-k_3-k_4)={\sqrt{2}\over |\sin\theta|Q^3}
\prod_{i=2}^4\delta(x_1-x_i)\,\delta(l_1+l_2-l_3-l_4),
\]
which indicates that the 4 constituents which enter or leave each hard
scattering carry the same light cone fraction ($x_i=x$ or $1-x$). 
Introducing the ``out of plane'' impact parameter $b$ through 
\[
2\pi\delta(l_i)=\int_{-\infty}^{+\infty}db\,e^{i(l_1+l_2-l_3-l_4).b},
\]
one may write the amplitude of the process~\cite{bot} as
\begin{equation}
\label{A0}
A(s,t)={\sqrt{2}Q\over 2\pi |\sin\theta|} 
\int_0^1dx\,(2\pi)^4H(\{xQv\})\,H'(\{\bar{x}Qv\})\Big|_{\{\alpha,\beta\}}
\int_{-\infty}^{+\infty}db\prod_{i=1}^4{\cal P}_{\alpha_i\beta_i}(x,b,Qv_i).
\end{equation}
where $\cal P$ is the amplitude
\begin{equation}
{\cal P}_{\alpha\beta}(x,b,Qv)=
\int {dy^- \over 2\pi}\,e^{ixQy^-}
<0|\,T\left(q_{\alpha}(y)\,\bar{q}_{\beta}(0)\right)|\pi(Qv)>
\Big|_{y=y^{-}v'+b\eta},
\end{equation}
\noindent
with Dirac indices $\alpha\beta$. Color indices are suppressed in 
Eq.~(\ref{A0}) and sums over repeated indices are understood. We consider 
unflavored quarks. The implementation of flavor is straightforward by setting 
to 0 some of the graphs we are going to consider. $H$ and $H'$ are Feynman 
amplitudes (a sum over allowed diagrams is understood) at the level of quarks 
computed with standard perturbative QCD vertices and internal propagators; 
they consist, at lowest order in the coupling constant, of one-gluon exchange 
or $q\bar{q}$ annihilation for each quark pair. Note that $\cal P$ and $H$ 
are not individually gauge invariant. If the short--distance 
$b\rightarrow 0$ limit is assumed, then the four general wave functions in 
Eq.~(\ref{pion}) can be reduced to one $A_\pi$--term, by taking 
the trace ${\cal P}_0$ of ${\cal P}_{\alpha\beta}$ with 
$Q/\!\!\!v'\gamma^5_{\alpha\beta}$. For the case of helicity conserving 
amplitudes, this selection was shown to be self--consistent by Botts and 
Sterman\cite{bot}. Then, the zeroth moment of ${\cal P}_0$ is related to the decay 
constant of the corresponding meson
\[
\int_0^1dx {\cal P}_0(x,b=0,p=Qv)=f_M
\]
where, {\it e.g.} for the pion, $f_\pi=133MeV$. This zero--distance quantity
contains no information on the interesting dependence on the transverse
variable $b$. 

Here we are concerned with the leading order description of helicity 
violating terms. Thus, we will consider $A_\pi$--type and $D_\pi$--type 
amplitudes on an equal footing, and make no a priori assumption that the 
region $b\rightarrow 0$ dominates. 


\subsection{Gauge Invariance}

The development so far has been sufficient to isolate the 
kinematic region of interest, which as we have already noted is 
characterized by finite separation between the participating quarks 
in the out-of scattering plane direction. The amplitude is thus a 
strong function of the spatial dependence of the wave function. The 
Bethe Salpeter wave function is a bilocal matrix element and is 
not gauge invariant. However, we will now discuss how gauge 
invariance of the description can be obtained. 

The key is in how the perturbation theory is re-arranged. In 
the Sterman and Botts factorization certain ``soft'' corrections are 
put into the wave functions, leaving other parts of Feynman 
diagrams to go into the hard scattering kernel. In dressing the wave 
function in this way, it is no longer a quark correlation (the Bethe 
Salpeter wave function), but the matrix elements of operators 
determined by the types of diagrams put in. The operators chosen 
in~\cite{bot} are path ordered exponentials (poe's), shown by 
Collins and Soper~\cite{col} to be the generators of eikonal approximations 
to the gluon attachments. The poe's are gauge covariant, leading to 
a gauge invariant amplitude.

This is partly forced by physics, and partly a convention. As a 
convention for the perturbation theory, subsequent diagrams must 
be evaluated with subtractions to avoid double-counting. More 
generally, any operator functional of the $A$ fields which transforms 
properly could serve in place of the poe's, and creating a different 
subtraction procedure. Let us extract what we can that is independent 
of convention.

Let the operator in the definition of the wave functions be 
called $U(A;x)$; we will call it a gauge-dressing operator. Under a 
gauge transformation at the position $x$, we require $U(A;x)$ to 
transform like an antiquark. Then products such as $U(A;x)\,\psi(x)$ 
are gauge invariant. That is, we have gauge invariant matrix elements 
to find a dressed quark
\[
\langle0|T(U(A;y)\psi(y)\,U^{\dagger}(A;x)\bar{\psi}(x)) |\pi(p)\rangle
\]

It is obvious that this requirement does not determine $U(A;x)$ 
uniquely, because one could always attach a factor which is gauge 
invariant without changing the gauge transformation properties. The 
particular choice of what to attach is a prescription, {\it i.e.} a 
definition of what parts of the amplitude will be put in the wave 
function and what in the hard scattering, and it cannot be determined 
by gauge invariance alone. However, due to gauge transformations one 
must attach some kind of gauge-dressing operator to have a well defined 
matrix elements.

\subsection{Path-Independent Dressing}

Although the standard way to do gauge dressing with the poe is 
path dependent, no path dependence generally need be associated 
with $U(A;x)$ and in particular the observable process does not 
determine or favor any path. This important point can be seen with 
an elementary example from QED, where the $U(1)$ gauge invariance is 
easier to control. The straightforward QED analogue of our process 
involves equal time (not light cone time) correlation functions in the 
gauge $A^0=0$. This gauge choice eliminates a mode, but there still 
remains a lack of definition of the coordinates due to time 
independent gauge transformations $\theta (x)$
\[
{\bf A}(x)\rightarrow {\bf A}'(x)={\bf A}(x)+ \nabla \theta(x),
\;\partial^0\theta=0
\]

This produces a change in the longitudinal modes. These modes are 
sometimes also called unphysical, a very unfortunate choice of 
terminology. In free space and in the absence 
of coupling a gauge theory has two transverse degrees of freedom 
and the third would be called unphysical.  However we are interested 
in the case that matter fields exist (and the non-Abelian coupling is 
turned on) in which case the third mode is real, but special inasmuch 
as being constrained in terms of the other variables by gauge 
invariance.  To see this, note that we can decompose into transverse
and longitudinal parts,
\begin{eqnarray}
{\bf A}   &=& {\bf A}_T + {\bf A}_L
\nonumber\\
	  &=& {\bf A}_T + \nabla\phi
\nonumber\\
\phi &=& \left[{1\over\nabla^2}\nabla\cdot{\bf A}\right](x)     
\hskip .25in \;\hbox{(transforming part)}
\nonumber\\
{\bf A}_T &=& {\bf A} -  \nabla\phi \hskip .75in\;\hbox{(invariant part).}
\end{eqnarray}

Since it is there for gauge transformations, the longitudinal part 
$\phi$ is not free to be varied in arbitrary dynamical ways, but 
must accompany the matter field in a prescribed, unique functional. 
At time $t=0$ this operator is 
\begin{equation}
U(A;x) = e^{i g\phi(x)}
\end{equation}
from which one can verify that under a gauge transformation,
\begin{eqnarray}
\phi(x)       &\rightarrow & \phi(x)+\theta(x)\nonumber\\
\psi(x)       &\rightarrow & e^{-ig\theta(x)}\psi(x) \nonumber\\
U(A; x)       &\rightarrow & e^{ig\theta(x)}U(A;x)   \nonumber\\
U(A;x)\psi(x) &\rightarrow & U(A;x)\psi(x)
\end{eqnarray}

We will call this the Coulomb dressing because it creates a classical 
Coulomb field around the matter particle, as the reader can check by 
calculation.  Since the Hamiltonian commutes with the gauge 
transformation operator once $A^0=0$ has been set, the time evolution 
will maintain the invariance of the combination $U(A;x)\psi(x)$.  
However, as noted already, one is not forced to accept this as the 
unique answer, but can opt for $U(A;x)\,f({\bf A}_T)$, which will 
transform in the same way for any $f({\bf A}_T)$.

The reader may still be curious to know the relation between 
Coulomb dressing and the poe approach. This can be very simply 
exhibited by noting that
\begin{equation}
\psi(y)e^{ig\phi(y)}e^{-ig\phi(x)}\bar{\psi}(x)
=\psi(y)e^{ig\int_x^yd{\bf z}.\nabla(\phi)({\bf z})}\bar{\psi}(x)
\end{equation}

This expression is still path independent.  This is the choice 
$f({\bf A}_T) =1$. A different choice is 
\begin{equation}
f({\bf A}_T) =e^{ig\int_x^yd{\bf z}.{\bf A}_T({\bf z})}
\end{equation}
in which case we have
\begin{equation}
U(A;y)U^{\dagger}(A; x)f({\bf A}_T) 
=e^{ig\int_x^yd{\bf z}.{\bf A}({\bf z})} 
\end{equation}
which is the standard path dependent poe.   Both procedures are 
equally acceptable, as far as satisfying gauge invariance, but the 
path ordered exponential can create a line of physical transverse 
gauge field particles between the charged matter fields, depending 
on how it is oriented.  Such a line of gluons, existing only along the 
chosen part, can be interpreted as an arbitrary model for the 
transverse gauge field inside the state of interest.  Similarly, if one 
boosts the Coulomb dressed definition, the boost also creates a 
blast of transverse gauge fields as seen by the ``equivalent photon'' 
approximation.  

In perturbation theory, the lowest order approximation to non-Abelian 
dressing is the Abelian case.  It is possible in the non-Abelian 
theory to write down expressions analogous to the Coulomb 
dressing but care must be used to keep track of the color indices.  It 
is equally valid to use poe's, which definitely transform properly, as 
building blocks to generate an infinite number of different ways to 
dress the quarks.  The different choices are not relevant for 
a leading order calculation to which we restrict this study. 

\subsection{Factorization}

The next crucial step is to elaborate a factorized form for the amplitude, 
whose prototype is Eq.~(\ref{A0}), regarding radiative corrections. 
Generalizing the results of \cite{bot} to the case of the
helicity--violating Dirac projections, a leading approximation to the soft
region rearranges these corrections to obtain the following 
expression
\begin{equation}
\label{Af}
A(s,t)={\sqrt{2}Q\over 2\pi |\sin\theta|}
\int_0^1dx\,H(\{xQv\})\,H'(\{\bar{x}Qv\})
\int_{-1/\Lambda}^{+1/\Lambda}db\,U(x,b,Q)
\left({\cal P}^{(0)}(x,b;1/|b|)\right)^4,
\end{equation}
\noindent
where $H$ and $H'$ are evaluated at respective scales $xQ$ and $\bar{x}Q$ 
which are assumed to be large.
Large logarithmic corrections to the process, with the coexistence of 
the two scales $Q$ and $1/b$, are resummed in $U$, in such a
way that ${\cal P}^{(0)}$ is now a soft object: it does not include
perturbative corrections harder than $1/b$. It is the non perturbative 
object necessary to connect short and long range physics, both present in 
hard hadronic processes. 

Expressions for the above quantities will be given in the phenomenological 
study, but let us specify here some of their qualitative features. When 
$b$ is small, smaller or of the order of $1/Q$, radiative corrections are 
also small and must be considered as perturbative corrections to the hard 
amplitudes $H$ or $H'$; in this regime, ${\cal P}^{(0)}$ is closely related 
to the 
distribution amplitude $\varphi(x;\mu)$ evaluated at the renormalization
scale $\mu=1/b$~\cite{bot}. Depending on the definition, the 
distribution amplitude also includes resummed logarithmic corrections 
from hadronic scale up to $\mu$~\cite{bro1,bro3}. When $b$ is large, this 
replacement is not legitimate but since $U(b)$ is a strong suppression 
factor in this region the exact value of ${\cal P}^{(0)}$ is irrelevant.

Endpoints in the $x$ integral where hard subprocesses would indeed become 
soft may look problematic. Both the distribution amplitude and wave function 
approaches used here become self--consistent because of the end--point 
zeroes, e.g. $\varphi(x\rightarrow 0)\sim x^k$, where $k>0$ which should 
occur independently of spin projection. 



\section{Contribution from non-zero orbital momentum components of the wave 
function: $\pi\pi\rightarrow\pi\pi$}

Before analyzing helicity violating processes let us examine the leading
contribution from the various components in Eq.~(\ref{pion})
to a standard helicity conserving reaction such as $\pi\pi\rightarrow\pi\pi$.

\subsection{Computation of hard amplitudes}

In their study, Botts and Sterman were interested in identifying the asymptotic 
behavior of the amplitude $A(s,t)$ ($s\rightarrow+\infty, {t\over s}$ fixed).
Asymptotically, the Sudakov mechanism contained in $U(b)$ results in a 
suppression of large $b$ region in the integral of Eq.~(\ref{Af}). In this
limit one can forget about tensorial components of 
${\cal P}_{\alpha\beta}(x,b,Qv)\propto (\cdots b\!\!\!/\cdots)_{\alpha\beta}$ 
and only the projection of $P_{\alpha\beta}(x,b,Qv)$ onto the particular tensor 
${1\over 4}\gamma_5v\!\!\!/|_{\alpha\beta}$ survives.

In the intermediate $Q^2$ regime, configurations of the $q\bar{q}$ pair sitting 
in a light meson with transverse separation smaller than the meson charge 
radius are not strongly affected by the Sudakov mechanism~\cite{jak}. As 
anticipated in section~2, $m=1$ components of the wave function which form 
large invariants in $H$ (as large as the s-wave term) may give sizeable 
corrections to the interaction amplitude between pure s-wave states. For 
contributions with leading power behavior 
in the pseudoscalar case, we must keep the tensorial decomposition
\begin{equation}
{\cal P}_{\alpha\beta}(x,b,Qv)=
{1\over 4}\gamma_5\left\{{\cal P}_0(x,b,Qv)v\!\!\!/+
{\cal P}_1(x,b,Qv)[v\!\!\!/,b\!\!\!/]\right\}_{\alpha\beta}.
\end{equation}

We now explore the calculation with this assumption. To begin with, one forms 
the projection, denoted as $t$ and $t'$, of the hard amplitudes $H$ and $H'$ 
on the relevant various Dirac tensors coming from the wavefunction 
decomposition. We follow 
Botts and Sterman's classification of graphs, with three fermionic flows for 
$H$ and two gluonic channels each (see Fig.~3),
\[
\begin{array}{ccccc}
f & H(M_1M_2\rightarrow M_3M_4) & \hbox{gluon} & c^f_{1,\{a_i\}} 
& c^f_{2,\{a_i\}}\\
1 & 1\bar{2}\rightarrow\bar{3}4 &u,\;s & \delta_{a_1a_2}\delta_{a_3a_4}
& \delta_{a_1a_4}\delta_{a_2a_3} \\
2 & 1\bar{2}\rightarrow3\bar{4} & t,\;s & \delta_{a_1a_2}\delta_{a_3a_4}
& \delta_{a_1a_3}\delta_{a_2a_4} \\
3 & 12\rightarrow34             & t,\;u & \delta_{a_1a_4}\delta_{a_2a_3}
& \delta_{a_1a_3}\delta_{a_2a_4}\\
\end{array}
\]
Color flow in this problem is simplified by noting that one gluon 
exchange between two quark lines gives a color tensor
\[
\sum_cT^c_{ij}T^c_{kl}={1\over 2}\left(\delta_{il}\delta_{kj}
-{1\over 3}\delta_{ij}\delta_{kl}\right),
\]
which we may reexpress with the color tensors listed above, in the form
\[
\left[t\,\hbox{or}\,u1 \right]_{\{a_i\}}
=\sum_{I=1}^2C_Ic^f_{I\{a_i\}},\;
\left[s\,\hbox{or}\,u3 \right]_{\{a_i\}}=\sum_{I=1}^2\tilde{C}_Ic^f_{I\{a_i\}},
\]
with
\[
C_1=\tilde{C}_2={1\over 2},\;C_2=\tilde{C}_1=-{1\over 6}.
\]

With this notation, one finds for the hard amplitude $a_0$, containing no 
$b$ factor,
\begin{eqnarray}
\left(t_It_J\right)_0^{(1)}&=&
 C_IC_J                                 {s^2+t^2 \over u^2}
+(\tilde{C}_IC_J+C_I\tilde{C}_J)        {t^2 \over su}
+\tilde{C}_I\tilde{C}_J                 {t^2+u^2 \over s^2};
\nonumber\\
\label{a0}
\left(t_It_J\right)_0^{(2)}&=&
 C_IC_J                                 {s^2+u^2 \over t^2}
+(\tilde{C}_IC_J+C_I\tilde{C}_J)        {u^2 \over st}
+\tilde{C}_I\tilde{C}_J                 {t^2+u^2 \over s^2};
\\
\left(t_It_J\right)_0^{(3)}&=&
 C_IC_J                                 {s^2+u^2 \over t^2}
+(\tilde{C}_IC_J+C_I\tilde{C}_J)        {s^2 \over tu}
+\tilde{C}_I\tilde{C}_J                 {s^2+t^2 \over u^2};
\nonumber
\end{eqnarray}
times an overall common factor
\[
\left({\pi\over 6}\right)^4{32g^4\over x^2\bar{x}^2s^2},
\]
where $g$ is the QCD coupling constant which appears in Feynman rules. 
We have already indicated that the whole amplitude Eq.~(\ref{Af}) 
can be properly defined regarding
renormalization and factorization, so that $g^4$ stands for 
$(4\pi)^2 \alpha_S(xQ) \alpha_S(\bar{x}Q)$.

There is no $b^1$ or $b^3$ terms, due to the odd number of $\gamma$
matrices; this is a consequence of chiral symmetry. 
The second term is therefore a hard amplitude $a_2$, containing 
$b^2$, which is found to be 
\begin{eqnarray}
\left(t_It_J\right)_2^{(1)}&=&
 C_IC_J                                 {2st \over u^2}
+(\tilde{C}_IC_J+C_I\tilde{C}_J)        {su-t^2 \over su}
+\tilde{C}_I\tilde{C}_J                 {2tu \over s^2},
\nonumber\\
\left(t_It_J\right)_2^{(2)}&=&
 C_IC_J                                 {2su \over t^2}
+(\tilde{C}_IC_J+C_I\tilde{C}_J)        {st-u^2 \over st}
+\tilde{C}_I\tilde{C}_J                 {2tu \over s^2},
\\
\left(t_It_J\right)_2^{(3)}&=&
 C_IC_J                                 {2su \over t^2}
+(\tilde{C}_IC_J+C_I\tilde{C}_J)        {tu-s^2 \over tu}
+\tilde{C}_I\tilde{C}_J                 {2st \over u^2},
\nonumber
\end{eqnarray}
\noindent
with a common factor
\[
-\left({\pi\over 6}\right)^4{256g^4\over x^2\bar{x}^2s^2}b^2,
\]
(here $b$ is a distance so that $b^2\ge 0$) 
and the third one a $b^4$ hard amplitude, $a_4$, which is the combination of 
Eq.~(\ref{a0}), but with an overall factor
\[
\left({\pi\over 6}\right)^4{512g^4\over x^2\bar{x}^2s^2}b^4.
\]

A check of the above expressions or a possibility to reduce the number
of graphs one has to compute is provided by symmetries under meson exchange;
starting from the expression one gets with two $u$-gluon exchange and fermionic 
flow $f=1$, see Fig.3, which we label $uu1$, one can generate
\[
\begin{array}{cccc}
\hbox{Exchange}		& \hbox{ channel}	& \hbox{ kinematic}	
& \hbox{ color}\\
2 \leftrightarrow 4	& ss1		& u \leftrightarrow s 
& C \leftrightarrow \tilde{C}\\
3 \leftrightarrow 4	& tt2		& u \leftrightarrow t 
& C \leftrightarrow C\\
2 \leftrightarrow 3	& uu3		& s \leftrightarrow t 
& C \leftrightarrow \tilde{C}.\\
\end{array}
\]
The reader will easily find the channels obtained from another starting point, 
say $us1$, and the combination of exchanges needed to determine graphs which 
do not appear in the above array, thus completing the whole amplitude.

\subsection{Asymptotic behavior}

We are now ready to evaluate the integral defined in Eq.~(\ref{Af}) with the 
above hard amplitudes. In a first step, we approximate the Sudakov factor 
by its dominant expression at large $Q$~\cite{bot}
\begin{equation}
\label{Ull}
U(x,b,Q) \approx \exp\left[-c\ln {xQ\over \Lambda}
(-\ln u(xQ,b)-1+u(xQ,b))\right]\,\exp\,[x\leftrightarrow \bar{x}],
\end{equation}
\noindent
with
\[ 
u(xQ,b)=\left(-{\ln b\Lambda\over\ln xQ/\Lambda}\right)
\]
and $\displaystyle c=4\,{4 \over 3}\,{2 \over 11-2n_f/3}=32/27$ for $n_f=3$. 
We have introduced the variable $u(x,b)$ which turns out to be the relevant 
one to describe the Sudakov unsuppressed region in the $(b,Q)$--plane: for 
$u(xQ,b)=1$ there is no suppression from the first exponential in 
Eq.~\ref{Ull}; as soon as $u$ departs from unity, one gets rapidly a strong
suppression due to the large $\ln xQ$ factor.
Eq.~(\ref{Ull}) is likely to be valid for any Dirac projection because 
the leading logs factor away independent of spin.
In this approximation, $U$ is a scalar in color space and one easily performs 
the color traces, with
\[
c_{I\{a\}}c_{J\{b\}}\prod_{i=1}^4\delta_{a_ib_i}
=3\left( \begin{array}{cc} 3 & 1 \\ 1 & 3 \end{array} \right),
\]
\[
C_IC_J={1\over 36}
\left( \begin{array}{cc} 9 & -3 \\ -3 & 1 \end{array} \right),\;
C_I\tilde{C}_J={1\over 36}
\left( \begin{array}{cc} -3 & 9 \\ 1 & -3 \end{array} \right),\;
\]
\[
\tilde{C}_I\tilde{C}_J={1\over 36}
\left( \begin{array}{cc} 1 & -3 \\ -3 & 9 \end{array} \right),
\]
and obtains the following hard amplitudes, labeled by the power of $b$ 
entering:
\begin{eqnarray}
a_0&=&\left({\pi\over 6}\right)^4{256g^4\over x^2\bar{x}^2s^2}
{s^4(s^2-3tu)+t^2u^2(s^2-tu)\over s^2t^2u^2};\nonumber\\
a_2&=&b^2\,\left({\pi\over 6}\right)^4{2048g^4\over x^2\bar{x}^2s^2}
{s^4(s^2-3tu)-t^3u^3\over s^2t^2u^2};\\
a_4&=&b^4\,\left({\pi\over 6}\right)^4{2048g^4\over x^2\bar{x}^2s^2} 
{s^4(s^2-3tu)+t^2u^2(s^2-tu)\over s^2t^2u^2};\nonumber
\end{eqnarray}
which at 90$^{\circ}$ is
\[
a_0=19\,\left({\pi\over 6}\right)^4{64g^4\over x^2\bar{x}^2s^2};\;
a_2=120 b^2\,\left({\pi\over 6}\right)^4{64g^4\over x^2\bar{x}^2s^2};
\]
\[
a_4=304b^4\,\left({\pi\over 6}\right)^4{64g^4\over x^2\bar{x}^2s^2}. 
\]


For the remaining soft wave function ${\cal P}_0$, one can adopt two different 
models. First, one can approximate ${\cal P}_0$ to its value at small 
$b$ (see~\cite{bot} and the argument in section 3.1)
\begin{equation}
{\cal P}_0(x,b;\mu) \approx \varphi (x;\mu),
\end{equation} 
with
\begin{eqnarray}
\label{phias}
\varphi_{\rm as}(x)&=&6 f_{\pi} x (1-x),\\
\label{phicz}
\varphi_{\rm cz}(x;\mu\sim 500MeV)&=&5(2x-1)^2\varphi_{\rm as}(x),
\end{eqnarray}
which are standard choices for pion distributions. The asymptotic form 
Eq.~(\ref{phias}) is derived in~\cite{bro1,bro3}; it has no evolution with 
$\mu$ and is indeed the limit 
as $\mu\rightarrow\infty$ of all distributions. Because evolution with $\mu$
is slow, the effective distribution at non asymptotic regime may be very 
different from $\varphi_{as}$. Chernyak and Zhitnisky~\cite{che} have built from
QCD sum rules the above cz-form Eq.~(\ref{phicz}), which evolves with 
the scale, but with  quite small effects on any computation when one does not 
examine a huge interval of energy. We will ignore these effects in the 
following.

The existence of the additional component ${\cal P}_1$ complicates the problem.
Next to nothing is known about it, but a reasonable ansatz is 
to adopt a form similar to ${\cal P}_0$. One notices that ${\cal P}_0$
has a mass dimension which appears to be set by $f_{\pi}$. In the case of
${\cal P}_1$, the dimension is a squared mass and we do not know 
what the normalization constant has to be. We will adjust the normalization in
the way described below, and assume the same 
$x$--dependence for ${\cal P}_0$ as ${\cal P}_1$.

For each hard part $a_i$, freezing the coupling at the scale $Q$ to simplify
the study, we get through Eq.~(\ref{Af}) a value $A_i(Q,90^{\circ})$ and 
perform the ratios $R_i=A_i/A_0$. 

To get the asymptotic behavior, we analytically evaluate the $b$-integral  
\[\int_0^{\Lambda^{-1}}db\,b^n U(b,x,Q)\]
with a saddle point approximation; using the change of variable 
$u=-\ln b/\ln \sqrt{x\bar{x}} Q$.
One has a maximum of the integrand at $u_0={2c \over 2c+n+1}$ and finds
\begin{equation}
\int_0^{\Lambda^{-1}}db\,b^n U(b,x,Q)\approx
u_0\,\sqrt{\pi \ln Q \over c} (x\bar{x} Q^2)^{c \ln u_0}.
\end{equation}
Performing the $x$-integration 
\[
I(\varphi,n)=\int_0^1dx \varphi(x)^4\,(x(1-x))
^{-2+c\ln {\scriptstyle 2c\over \scriptstyle 2c+n+1}},
\]
an asymptotic expression is found for the ratio
\begin{equation}
\label{Ras}
R_n(Q)={2c+1 \over 2c+n+1} 
\left({Q\over \Lambda}\right)
^{2c\ln {\scriptstyle 2c+1\over \scriptstyle 2c+n+1}}
{I(\varphi,n)\over I(\varphi,0)},
\end{equation}
from which one deduces
\[
R_2 \propto Q^{-1.10},\;R_4 \propto Q^{-1.85}.
\]
Roughly speaking, each power of $b$ is suppressed by $1/\sqrt{Q}$ (not $1/Q$).


\subsection{Intermediate behavior}

At accessible energies, we expect deviations from the result 
given in Eq.~(\ref{Ras}). This comes first from an eventual 
failure of the saddle point approximation with too small $\ln Q$. 
We thus have numerically evaluated the amplitude Eq.~(\ref{Af}) with $U$ 
given by Eq.~(\ref{Ull}). Results for our computation of the ratio of 
amplitudes are displayed in Fig.~4 with cz distribution (solid line). 
We get similar results for the asymptotic distribution amplitude (Fig.~5). 
To fix the normalization, we choose here and in the following to set 
arbitrarily the ratios to 1 at $\sqrt{s}=2$GeV. We observe that $R_2$ 
decreases by a factor around 7 from $\sqrt{s}=2$GeV to 20GeV. This is a much 
milder suppression than the naively expected $1/Q^2$ factor (dotted curve).
$R_4$ drops more drastically by a factor around 20 in the same energy 
interval. A numerical study shows that logarithmic
corrections ignored within the saddle point approximation are not negligible 
in the accessible range of energies. At larger values of the energy, 
$\sqrt{s}>20$GeV, the approximated result of Eq.~(\ref{Ras}) becomes accurate.

Secondly, Eq.~(\ref{Ull}) should be supplemented with non leading terms 
which are known in the s-wave case~\cite{bot} but presumably different in 
the p-wave case. The neglected logs in the expression of $U$ Eq.~(\ref{Ull}) 
may modify soomewhat the ratio over some intermediate range of energy. 
To modelize such an effect, we add a simple $x,b$-independent term like
\[
\exp\left(K\ln\ln {Q\over\Lambda}\right)
\]
in the expression of $R_2$ with $K$ some constant. The 
ratio modified by such a factor is shown in Fig.~4 as a shaded area limited 
by the curves corresponding to $K=1$ and $K=-1$; this measures in some 
way the theoretical uncertainty on p-wave contribution suppression. 
Further theoretical progress in the computation of these $\ln\ln Q$ terms 
will allow to get rid of this uncertainty but we do not tackle this task here.

A third effect may come from the intrinsic transverse dependence of the 
wave function. While the replacement of ${\cal P}^{(0)}(x,b;1/b)$ by 
$\varphi(x;1/b)$ discussed in section~3.1 is reasonable at large 
$Q$~($>10$GeV), it is more questionnable at intermediate
values where long range physics may be accounted for by including
some intrinsic $b$-dependence~\cite{jak} as
\begin{equation}
{\cal P}_0(x,b)= 4 \pi {\cal N} \varphi (x) 
\exp\left(-{\alpha^2\over x\bar{x}}-{x\bar{x}b^2\over 4\beta^2}\right).
\end{equation}
with parameters $\alpha^2=.096$, $\beta^2=.88$GeV$^2$ and 
${\cal N}=1.68$ for the asymptotic distribution amplitude. The results are 
modified as depicted in Fig.~5 where the curves from this wave function and 
from the asymptotic distribution are shown for comparison. Furthermore, we 
notice that the phenomenology may also be modified considerably by allowing 
variations of different distribution form, that is $\varphi_{\rm as}$ for 
${\cal P}_0$ and $\varphi_{\rm cz}$ for ${\cal P}_1$ or the reverse choice.

These three effects show that the power-like decrement Eq.~(\ref{Ras})
of the ratio is diluted at intermediate energies and consequently the 
amplitudes $A_0$ and $A_2$ are likely to compete over a rather large interval 
of $s$, say 1GeV$^2$--100GeV$^2$.


\section{Helicity violating processes: 
$\pi\pi\rightarrow\rho\rho$ ($h_3+h_4\neq 0$)}

\subsection{Double helicity flip}

To begin with, let us find the possible tensorial decomposition for
the quark antiquark wavefunctions of the $\rho$ meson. One specifies a 
one particle state by the momentum $p_i$ ($i=3,4$), which with the notation 
of section~3, 
one can write: $p_i=Qv_i+{m^2\over 2Q}v_i'$ 
(m is the $\rho$ mass which we do not 
neglect for the moment), and by the helicity $h_i\in \{1,0,-1\}$. The 3 
helicity 
states are described in a covariant way with the help of the 3 vectors 
$\varepsilon_h^{\mu}(p_i)$, satisfying $\varepsilon_h.p_i=0$ and 
$\varepsilon_h.\varepsilon_{h'}^*=-\delta_{hh'}$. 
For $\rho(p_3)$, $(\vec{p}_3,\vec{\xi}_3,\vec{\eta})$ is a direct trihedral, 
and we choose
\begin{eqnarray}
(h=+1)\hspace{3cm} \varepsilon_R&=&-{1\over \sqrt{2}} (\xi_3+i\eta), \nonumber
\\
(h=\,\ \ 0)\hspace{3cm} \,\varepsilon_0&=&{Q\over m} v_3-{m\over 2Q}v_3', 
\\
(h=-1)\hspace{3cm} \varepsilon_L&=&{1\over \sqrt{2}} (\xi_3-i\eta). \nonumber
\end{eqnarray}
A consistent choice for $\rho(p_4)$ is
\begin{eqnarray}
(h=+1)\hspace{3cm} \varepsilon_R&=&{1\over \sqrt{2}} (\xi_3-i\eta), \nonumber
\\
(h=\,\ \ 0)\hspace{3cm} \,\varepsilon_0&=&{Q\over m} v_4-{m\over 2Q}v_4', 
\\
(h=-1)\hspace{3cm} \varepsilon_L&=&-{1\over \sqrt{2}} (\xi_3+i\eta). \nonumber
\end{eqnarray}

Thus, demanding a parity $-$ state, the Bethe-Salpeter amplitude has the most
general Dirac-matrices expansion
\begin{eqnarray}
&&{\cal P}(x,b;p,h)=\\
&&{\cal P}_0\,[/\!\!\!\varepsilon_h,/\!\!\!p]+{\cal P}'_0\,/\!\!\!\varepsilon_h+
{\cal P}_1\,\varepsilon_h.b\,/\!\!\!p+{\cal P}'_1\,\varepsilon_h.b+
\tilde{\cal P}_1\,[/\!\!\!\varepsilon_h,/\!\!\!p]\,/\!\!\!b+
\tilde{\cal P}'_1\,[/\!\!\!\varepsilon_h,/\!\!\!b]+
{\cal P}_2\,\varepsilon_h.b\,[/\!\!\!p,/\!\!\!b]+
{\cal P}'_2\,\varepsilon_h.b\,/\!\!\!b.\nonumber
\end{eqnarray}

One can then extract the relevant components for the study of independent
scattering processes. That is, we isolate dominant high-energy tensors, 
those which contain one power of the large scale $Q$ and we restrict to
$b_T^{\mu}$ ({\it i.e.} $b\eta^{\mu}$ in the notation of section~3.1). 
We get for a longitudinally polarized $\rho$
\begin{equation}
{\cal P}(x,b_T;p,h=0)={Q\over m}\left({\cal P}'_0 \, /\!\!\!v+
\tilde{\cal P}'_1\,[/\!\!\!v,/\!\!\!b_T]\right),
\end{equation}
and for a transversally polarized one
\begin{equation}
{\cal P}(x,b_T;p,|h|=1)=Q\left({\cal P}_0\,[/\!\!\!\varepsilon_h,/\!\!\!v]+
{\cal P}_1\,\varepsilon_h.b_T\,/\!\!\!v+
\tilde{\cal P}_1\,[/\!\!\!\varepsilon_h,/\!\!\!v]\,/\!\!\!b_T+
{\cal P}_2\,\varepsilon_h.b_T\,[/\!\!\!v,/\!\!\!b_T]\right).
\end{equation}

Before going through the computation of an helicity violating process, let us 
exhibit what we call the {\it double-flip} rule. It follows from the 
property of each dominant component that each power of $b$ changes the 
chirality (which is $+$ if the component anticommutes with $\gamma_5$ and 
$-$ if it commutes). With vector gluon couplings, it follows that if the 
process is allowed by hadron helicity conservation, chiral symmetry requires 
that the $b$-components occur in pairs.\footnote{The power suppression 
for amplitudes with extra transverse gluon in initial 
or final Fock states follows from the same consideration. By power counting 
one extra gluon embedded into one of the 2 hard scatterings is down by 
$1/Q$ with respect to the short distance contribution from valence states. 
However the chirality of the component is the same as the p-wave we are 
considering (that is opposite to the s-wave). The above argument then applies 
to ensure that when p-waves and one-gluon components are likely to compete, 
at least 2 such components are necessary: from section~4, one counts 
$Q^{-0.55}$ for one p-wave and $Q^{-1}$ for one extra gluon; this shows 
that considering one extra gluon in the Fock state is indeed subdominant with
respect to considering a p-wave state.} 
We have already encountered this property in the $\pi\pi$ elastic scattering.
Furthermore, for an helicity 0-state, in both pseudoscalar and 
vector case, the chirality of the s-wave term is positive, whereas for an 
helicity odd state, it is negative. Then with chiral symmetry and because the 
total power of $b$ has to be even, one concludes that the number of helicity-0 
states has to be even too. For a 2 to 2 process, only those processes that 
violate the helicity conservation rule by two units are expected to be non-zero
in the present framework and hence important at large energy.


An example of interaction which satisfies neither the helicity 
conservation rule nor the double-flip rule is 
\[
\pi\pi\rightarrow\rho_0\rho_T;
\]
such an amplitude is thus vanishing by power counting.

We then turn to the study of an helicity violating process which, however, 
satisfies the double-flip rule, namely $\pi\pi\rightarrow\rho_R\rho_R$.

\subsection{$\pi\pi\rightarrow\rho_R\rho_R$}

Let us first verify of the vanishing of the hard amplitude using the 
s-wave components of the external mesons. For this purpose, it is instructive 
to examine the connection between quark helicities and the s-wave Dirac tensors
we have used until now. This is accomplished in the following way. One 
considers the free massless spinors of a quark and an antiquark moving in the 
same direction, the quark having a momentum $xp$ and the antiquark a momentum 
$\bar{x}p$, so that the compound system has a momentum $p$. Then one 
constructs the four possible helicity states of the system with solutions of 
the Dirac equation and finds
\begin{eqnarray}
(\pi\,)\hspace{1cm}
{1\over \sqrt{2}}\left(
u_{\alpha}(xp,\uparrow)\bar{v}_{\beta}(\bar{x}p,\downarrow)-
u_{\alpha}(xp,\downarrow)\bar{v}_{\beta}(\bar{x}p,\uparrow)\right)
&=&
-\sqrt{x\bar{x}\over 2}\,\gamma_5p\!\!\!/|_{\alpha\beta} \nonumber
\\
(\rho_0)\hspace{1cm}
{1\over \sqrt{2}}\left(
u_{\alpha}(xp,\uparrow)\bar{v}_{\beta}(\bar{x}p,\downarrow)+
u_{\alpha}(xp,\downarrow)\bar{v}_{\beta}(\bar{x}p,\uparrow)\right)
&=&
-\sqrt{x\bar{x}\over 2}\,p\!\!\!/|_{\alpha\beta} \nonumber
\\
(\rho_R)\hspace{53mm}
u_{\alpha}(xp,\uparrow)\bar{v}_{\beta}(\bar{x}p,\uparrow)&=&
\sqrt{x\bar{x}\over 2}\,/\!\!\!\varepsilon_R p\!\!\!/|_{\alpha\beta} \nonumber
\\ 
(\rho_L)\hspace{53mm}
u_{\alpha}(xp,\downarrow)\bar{v}_{\beta}(\bar{x}p,\downarrow)&=&
\sqrt{x\bar{x}\over 2}\,/\!\!\!\varepsilon_L p\!\!\!/|_{\alpha\beta}, 
\end{eqnarray}

The helicity conservation rule is then easily verified with these combinations
of spinor when one chooses the chiral representation~\cite{far1}. In this 
representation, the 2 diagonal blocks of each $\gamma^{\mu}$ are equal to the
null 2$\times$2 matrix. A Feynman-graph fermion line, with vector (or axial) 
couplings and massless propagator, is an even number of $\gamma$--matrices
between two spinors:
\[
\psi'^{\dagger}(p',h')\gamma_0\gamma^{\mu_1}k\!\!\!/_1\cdots
k\!\!\!/_n\gamma^{\mu_{n+1}}\psi(p,h),
\]
Inclusions of $\gamma_5$, which is diagonal
in this representation, do not modify this property. Then, since to order $m/p$
the 
chirality of a spinor corresponds to the 
helicity of the state and, in this representation,
chiral eigenstate spinors have either their 2 first components or their two 
last equal to 0, states of different helicity always give a null product.

This property is algebraic and therefore independent of the representation 
chosen. However, it is more difficult to observe its effects in the trace
formalism described in section 2. Let us examine how it works in the case
of the reaction considered. Among the twelve graphs discussed in 
section~2, each $tt$ and $uu$ graph is 0 because they contain traces over 
the product of an odd number of $\gamma^{\mu}$ matrices. For the 8 remaining 
graphs, the sequence 
\[
/\!\!\!\varepsilon_{3R}v\!\!\!/_3\{\gamma^{\mu_1}\cdots\gamma^{\mu_{2n+1}}\}
/\!\!\!\varepsilon_{4R}v\!\!\!/_4\cdots
\]
occurs (the anticommutation to the left of every $\gamma_5$ is understood and 
does not modify the reasoning). The product of an odd number of 
$\gamma^{\mu}$ being a linear combination of $\gamma^{\mu}$ and 
$\gamma_5\gamma^{\mu}$, one is left with the evaluation of
\[
/\!\!\!\varepsilon_{R}\,v\!\!\!/\,\gamma^{\mu}\,
/\!\!\!\varepsilon_{L}\,v\!\!\!/'\cdots\;\hbox{ and}\;
/\!\!\!\varepsilon_{R}\,v\!\!\!/\,\gamma_5\gamma^{\mu}\,
/\!\!\!\varepsilon_{L}\,v\!\!\!/'\cdots,
\]
where $v_4=v'_3$ and $\varepsilon_{4R}=\varepsilon_{3L}$ have been used  
and the index 3 dropped. One can decompose each $\gamma^{\mu}$ onto
$(v\!\!\!/,\,v\!\!\!/',\,/\!\!\!\varepsilon_{R},\,/\!\!\!\varepsilon_{L})$,
which are such that their square is 0 and their anticommutation rules are
\[
\{/\!\!\!v,/\!\!\!\varepsilon_R\}=\{/\!\!\!v,/\!\!\!\varepsilon_L\}=
\{/\!\!\!v',/\!\!\!\varepsilon_R\}=\{/\!\!\!v',/\!\!\!\varepsilon_L\}=0,
\]
to conclude that all graphs effectively vanish.

Adding the contribution from one p-wave function does not give any 
contribution (even before the integration over $b$), 
because the total number of $\gamma^{\mu}$ 
is necessarily odd. The first non zero term is therefore a $b^2$ hard 
amplitude $M_2$ and the computation, within the approximation of section~4.2, 
leads to
\begin{eqnarray}
M_2\left(\pi\pi\rightarrow\rho_R\rho_R\right)&=&
\left({\pi\over 6}\right)^4{128g^4\over x^2\bar{x}^2t^2u^2} b^2\left\{
{16(3s^2-7tu)\over 3}{\cal P}_{1\pi}^2{\cal P}_{0\rho}^2
-{t^3u^3\over s^4}{\cal P}_{0\pi}^2{\cal P}_{1\rho}^2\right.
\nonumber\\
&+&8{t^3u^3\over s^4}{\cal P}_{0\pi}^2{\cal P}_{0\rho}{\cal P}_{2\rho}
-16(s^2-3tu){\cal P}_{0\pi}{\cal P}_{1\pi}{\cal P}_{0\rho}{\cal P}_{3\rho}
\nonumber\\
&+&4{t^3u^3\over s^4}{\cal P}_{0\pi}^2{\cal P}_{1\rho}{\cal P}_{3\rho}
+\left. 4(s^2-3tu+{t^2u^2\over s^2}-2{t^3u^3\over s^4})
{\cal P}_{0\pi}^2{\cal P}_{3\rho}^2\right\}.
\end{eqnarray}
\noindent
This amplitude $M_2$ has to be supplemented with $U$ (Eq.~(\ref{Ull})) and 
integrated over $b$ and $x$.

Even though this combination involves several unknown objects, one notices
that the angular dependence varies from one component to another and is rather 
different from the one obtained in $\pi\pi$ elastic scattering. Therefore, it 
may be possible to analyze the contribution to helicity violation processes 
from different wave functions and use this information to deduce properties of
the wave functions.

The numerical study of section~4 can then be used to understand the energy 
dependence of double helicity violating processes at accessible energies. 
As explicitly shown in Fig.~4a and 5, the naive $1/Q^2$ factor is replaced 
by a milder suppression, and this is primarily due to the specificity of 
the independent scattering mechanism supplemented by Sudakov effects. Even 
at very large energies the $Q^{-1.10}$ ratio of Eq.~(\ref{Ras}) looks quite 
weak a suppression.  

\section{Realistic processes and experimental outlook}

Studying meson-meson scattering is an unrealistic simplifying 
assumption. Including baryons is a necessary but quite intricate 
further step, because of the high number of Feynman diagrams and 
internal degrees of freedom to be integrated over when 
more valence quarks are involved. We can however still draw some 
conclusions from our analysis, leaving to future work a comprehensive 
study. The mechanism we have explored occurs in several experimentally 
accessible circumstances.  Indeed, there is a host of reactions 
involving hadronic helicity violation from which we could learn about 
the interface of perturbative and non-perturbative QCD.  


The helicity density matrix of the $\rho$ meson produced in 
$\pi p\rightarrow\rho p$ at 90$^\circ$ is a nice measure of 
helicity violating components. Experimental data~\cite{hep} yield 
$\rho_{1-1}=0.32\pm0.10$, at $s=20.8$GeV$^2$, $\theta_{\rm CM}=90^\circ$,
for the non-diagonal helicity violating matrix element. Without 
entering a detailed phenomenological analysis, we may use the 
results of section~5 through the following line of reasoning. Assuming 
that the presence of the third valence quark, which is not subject to 
a third independent scattering, does not alter much the results, one may view 
$\rho_{1-1}$ as coming from the interference of an helicity conserving  
amplitude like $\pi\pi\rightarrow\rho_L\rho_R$ with a double helicity flip 
amplitude like $\pi\pi\rightarrow\rho_R\rho_R$. We then get a mild energy 
dependence of this matrix element {\it i.e.} $Q^{-1.10}$ (Eq.~(\ref{Ras})) 
at asymptotic energies or as shown in Fig.~4a and Fig.~5 at accessible 
energies. This is at variance with the picture emerging from the diquark 
model~\cite{ans}. Measuring the energy dependence of this effect is thus 
highly interesting.

The most well-known example of hadron helicity violation occurs in 
$pp\rightarrow pp$ scattering~\cite{cam}. Our demonstration of 
helicity violating contributions to meson-meson scattering 
has a bearing on this, because generalized meson scattering 
is embedded in the diagrams for proton scattering.  Without 
needing to make any dynamical assumption of ``diquarks'', 
the perturbative QCD diagrams for $pp\rightarrow pp$ scattering 
contain numerous diquark regions, convolved with 
scattering of an extra quark.  There is no known selection 
rule which would prevent the scattering of such a 
subprocess with helicity flip from causing helicity flip in pp 
elastic scattering.  This does not exhaust the possibilities, 
because there are other channels of momentum flow and 
color combinatorics which might have different 
interpretations.  The data for $pp\rightarrow pp$ also reveals large 
oscillations about power-law behavior, a second piece of 
evidence that the short distance picture is inadequate.  
Elsewhere~\cite{pir} we have identified these oscillations as a sign 
of independent scattering.  Given the theoretical~\cite{far} and 
experimental evidence, we therefore find no evidence 
that hadronic helicity conservation is a feature of perturbative 
QCD, and we believe that independent scattering is a main 
contender in explaining the observations.

Since reactions of baryons are extremely complicated, and next to nothing 
is known about the various wave functions in the proton, a productive 
approach to the question is to ask for experimental circumstances in which 
the general mechanism we have outlined could be tested without 
requiring too much detail. We believe that progress here 
will come from using nuclear targets, and studying the 
phenomenon of color transparency in hard (as opposed to 
diffractive) reactions. This program has been outlined 
elsewhere~\cite{ral2}; it suffices to mention here that suppression of 
large $b^2$ regions is expected in reactions of large nuclei. It 
follows that helicity conservation should be obtained in the 
same circumstances. Thus the mechanism we have outlined 
is testable. It is time to go beyond the short-distance physics 
used to establish QCD as the correct theory, and we believe 
that a multitude of phenomena involving spin, color 
transparency, and detailed hadron structure, will play a 
major role in the future.

\bigskip
\noindent {\bf Acknowledgments}:  This work was supported in part under 
DOE Grant Number DE FG02 85 ER 40214 and the 
Kansas Institute for Theoretical and Computational Science, and by the EEC 
program ``Human Capital and Mobility'', network ``Physics at high energy 
colliders'', contract CHRX--CT93--0357. 
We thank Stan Brodsky, Steve Heppelmann, Al Mueller, Jacques Soffer, 
George Sterman, and Joseph Zhitnitsky for useful discussions. 
Centre de Physique Th\'eorique is {\it unit\'e propre du CNRS}.

\newpage

\newpage

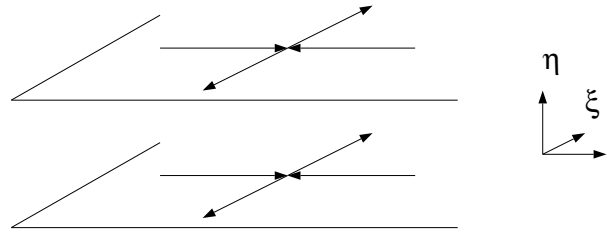
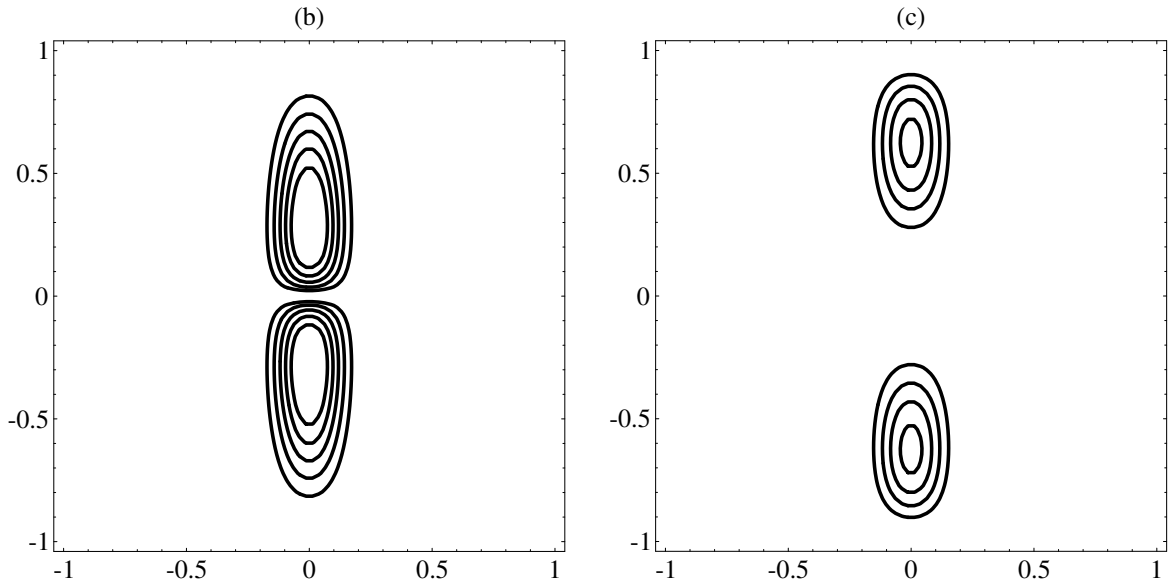
Figure 1: Coordinate space pictures of meson--meson independent
scattering. (a) Trajectories of quarks. In the scattering plane quarks approach
each other within a distance of order $1/Q$, while the transverse separation
of scattering planes (indicated in perspective view) is larger and set by the
wave function. (b) Contour map of real part of the product of $m=0$ leading
Sudakov wave functions (from \protect\cite{bot}) which give a model for the integrand of Eq.~(\protect\ref{A0}). No other soft wave function has been 
introduced. (c) Same wave function multiplied by a polynomial representing 
$m=2$. Contours in (b,c) show flattening in the $\xi$-direction of the 
effective wave function imposed by a Gaussian with 
$b_{\xi}^2<1/Q^2$ at $Q^2=$2 GeV$^2$; higher $Q^2$ increases 
the flattening. Units of $b_{\eta}$ and $b_{\xi}$ are fm.
\bigskip

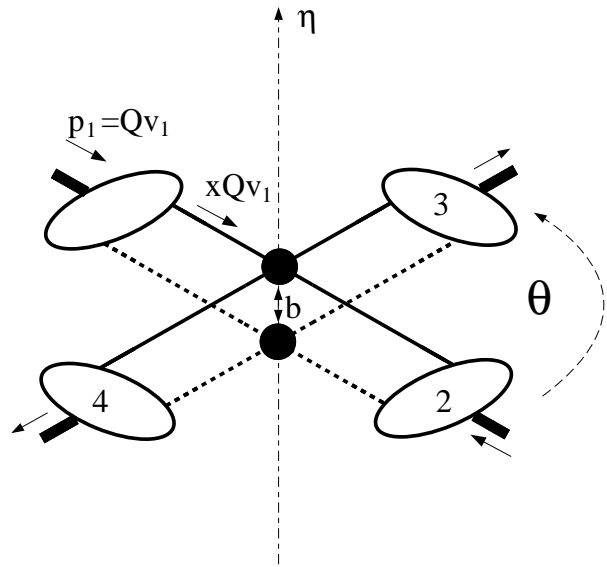
Figure 2: Kinematics of the independent scattering mechanism.
\bigskip

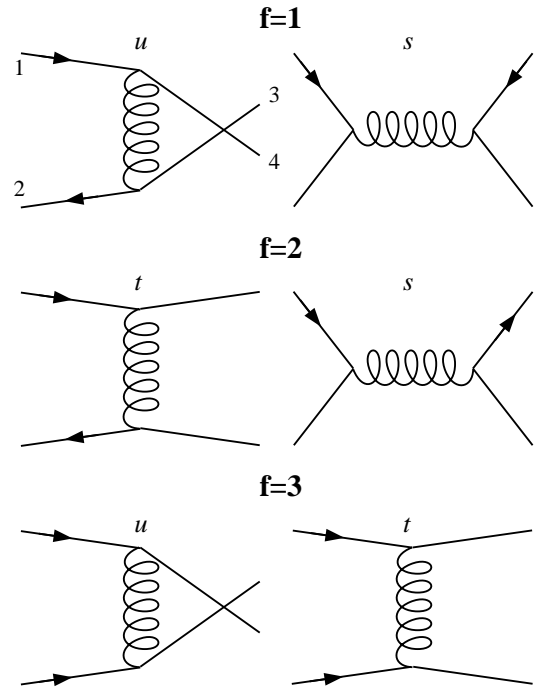
Figure 3: Feynman graphs for the lowest order hard amplitude $H$; for 
$H'$ reverse the arrows.
\bigskip

Figure 4: The energy dependence of the $R_2$ (a) and of the $R_4$ (b) ratios
with cz  distribution amplitude~\protect\cite{che} (thick lines) and their 
naive behaviors (thin lines), respectively $1/Q^2$ and $1/Q^4$. The shaded 
area in (a) indicates uncertainty from neglected logs. The ratios are 
normalized to 1 at $\protect\sqrt{s}=$2GeV.
\bigskip

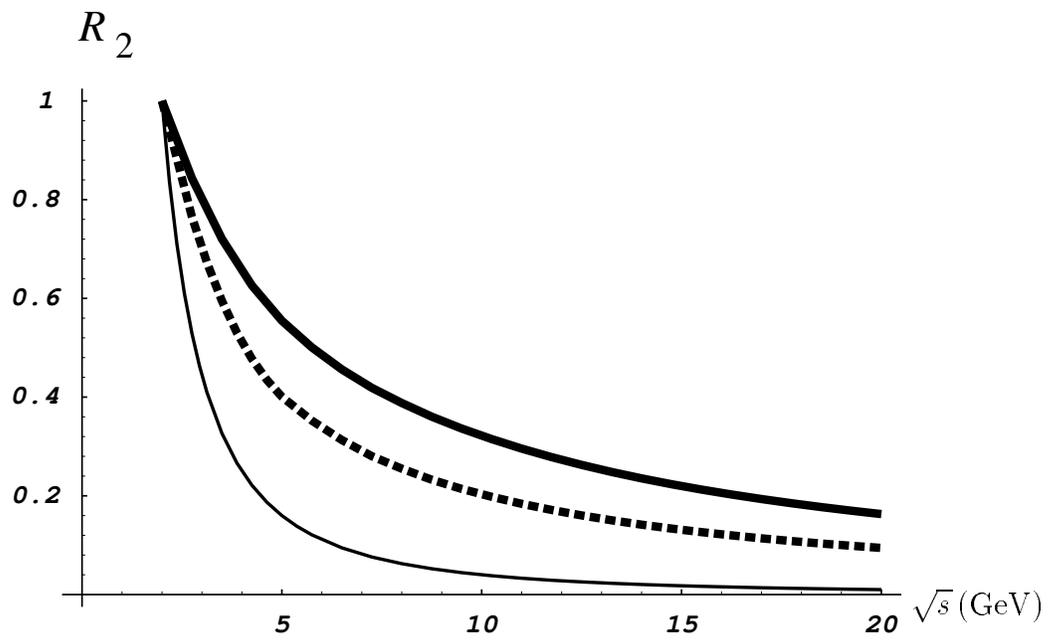
Figure 5: The energy dependence of the $R_2$ ratio calculated with the 
asymptotic distribution amplitude with (solid thick line) and without (dashed 
line) intrinsic $b$-dependence. The thin line is as in Fig.~4.
\bigskip

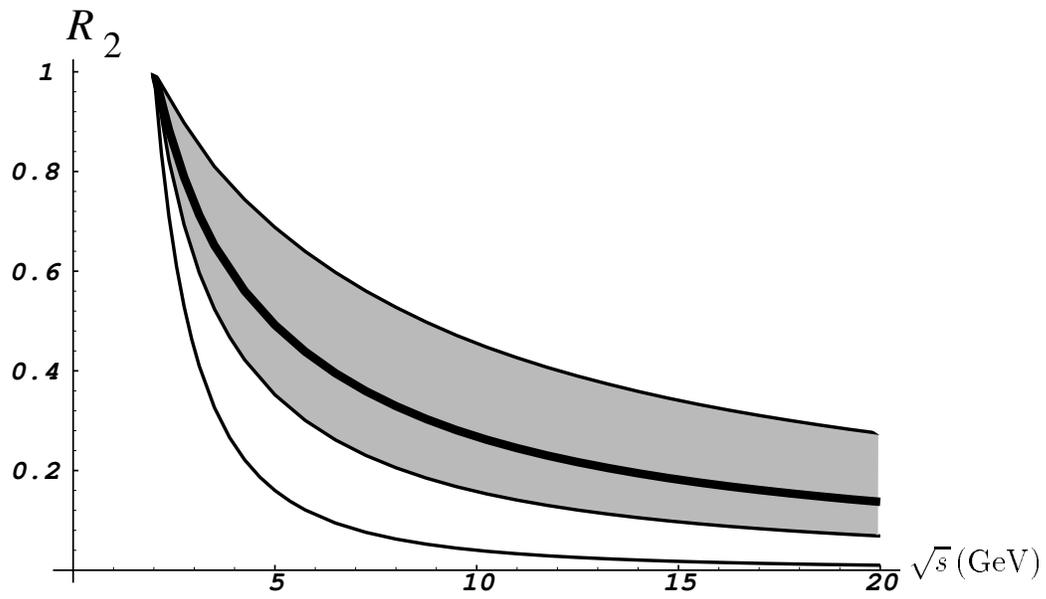
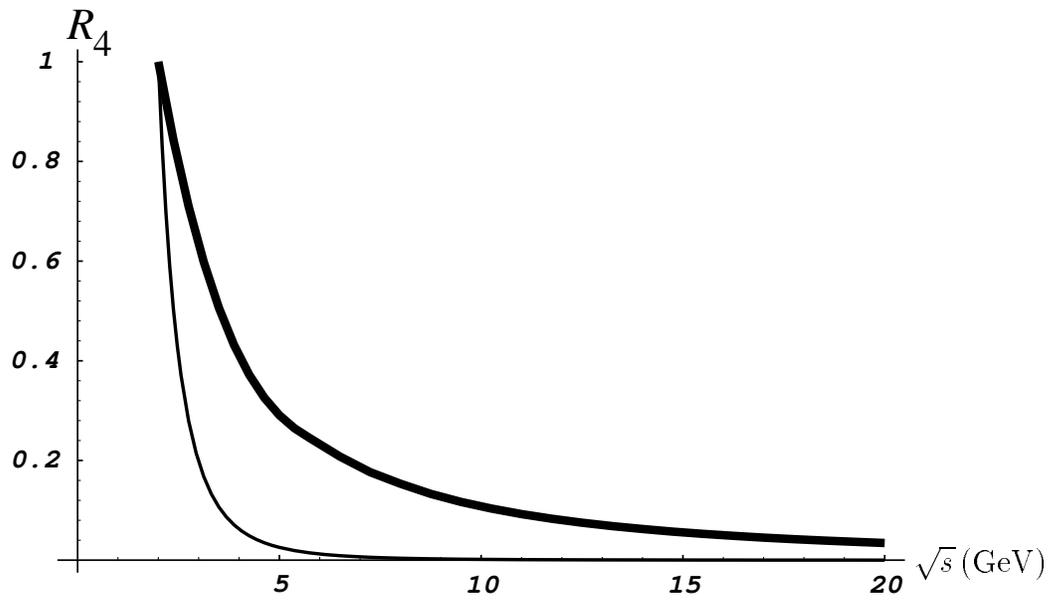
\end{document}